\documentclass[%
reprint,
superscriptaddress,
 amsmath,amssymb,
 aps,
]{revtex4-2}

\usepackage{siunitx}
\DeclareSIUnit{\belmilliwatt}{Bm}
\DeclareSIUnit{\dBm}{\deci\belmilliwatt}

\usepackage{graphicx}
\usepackage{dcolumn}
\usepackage{multirow}
\usepackage{tabularx}
\usepackage{ulem}
\usepackage{fancyhdr}
\usepackage{array}
\usepackage{booktabs}
\usepackage{color}
\usepackage{bm}
\usepackage{eurosym}
\usepackage{subfigure}
\usepackage{comment}
\usepackage{scrhack}
\usepackage{physics}
\usepackage{balance}
\usepackage{float}
\usepackage{flushend}
\usepackage{soul}
\usepackage{amsmath}
\usepackage[utf8]{inputenc}
\usepackage[T1]{fontenc}
\usepackage{mathptmx}
\usepackage{etoolbox}

\begin{document}

\preprint{APS/123-QED}
\title{Light-induced dynamic frequency shifting of microwave photons \\ in a superconducting electro-optic converter }%


\author{Yuntao Xu}
\affiliation{Department of Electrical Engineering, Yale University, New Haven,
Connecticut 06511, USA}
\author{Wei Fu}
\affiliation{Department of Electrical Engineering, Yale University, New Haven,
Connecticut 06511, USA}
\author{Yiyu Zhou}
\affiliation{Department of Electrical Engineering, Yale University, New Haven,
Connecticut 06511, USA}
\affiliation{Yale Quantum Institute, Yale University, New Haven,
Connecticut 06511, USA}
\author{Mingrui Xu}
\affiliation{Department of Electrical Engineering, Yale University, New Haven,
Connecticut 06511, USA}
\author{Mohan Shen}
\affiliation{Department of Electrical Engineering, Yale University, New Haven,
Connecticut 06511, USA}
\author{Ayed Al Sayem}
\affiliation{Department of Electrical Engineering, Yale University, New Haven,
Connecticut 06511, USA}
\author{Hong X. Tang}
\thanks{Corresponding author: hong.tang@yale.edu}     
\affiliation{Department of Electrical Engineering, Yale University, New Haven,
Connecticut 06511, USA}
\affiliation{Yale Quantum Institute, Yale University, New Haven,
Connecticut 06511, USA}

\date{\today}

\begin{abstract}
Hybrid superconducting-photonic microresonators are a promising platform for realizing microwave-to-optical transduction. However, the absorption of scattered photons by the superconductors leads to unintended microwave resonance frequency variation and linewidth broadening. Here, we experimentally study the dynamics of this effect and its impact on microwave-to-optics conversion in an integrated lithium niobate-superconductor hybrid resonator platform. We unveiled an adiabatic frequency shifting of the intracavity microwave photons induced by the fast photo-responses of the thin-film superconducting resonator. As a result, the temporal and spectral responses of electro-optics transduction are modified and well described by our theoretical model. This work provides important insights on the light-induced conversion dynamics which must be considered in future designs of hybrid superconducting-photonic system.   

\end{abstract}

\maketitle


\section{Introduction}

A hybrid superconducting-photonic transducer serves an important quantum interface for future quantum networks \cite{schoelkopf2008wiring,zhong2020proposal,PhysRevA.76.062323,cirac1997quantum,kimble2008quantum,monroe2014large} where the quantum information processed in microwave frequency qubits \cite{clarke2008superconducting,devoret2013superconducting} could be distributed by photonic circuits \cite{o2009photonic}. In various microwave-to-optical transduction schemes \cite{lauk2020perspectives,han2021microwave,lambert2020coherent}, such as cavity electro-optics (EO) \cite{fan2018superconducting,sahu2022quantum,McKenna:20,fu2020ground,Holzgrafe:20,xu2021bidirectional}, magneto-optics \cite{hisatomi2016bidirectional,zhu2020waveguide,bartholomew2020chip}, and piezo-optomechanics \cite{han2020cavity,forsch2020microwave,mirhosseini2020superconducting,shao2019microwave,jiang2020efficient}, an intense optical pump is required to uplift the conversion efficiency. 
Harnessing the Pockels effect, cavity EO systems are particularly attractive for achieving direct transduction between microwave photons and telecom optical photons without involving an intermediate excitation (phonon or magnon) whose thermal excitation may add noise to the full conversion chain \cite{tsang2010cavity,tsang2011cavity,javerzac2016chip}. Recently, a pulsed-pump microwave-to-optical conversion scheme has been developed to take advantage of a strong optical pump while reducing the overall thermal heating effect in the cryogenic environment \cite{fu2020ground,sahu2022quantum,forsch2020microwave,han2020cavity,mirhosseini2020superconducting}. In a bulky lithium niobate EO transducer, an impressive, near-unity cooperativity has been achieved by using watt-scale pump pulses \cite{sahu2022quantum}.

Moving onto an on-chip platform, the integrated cavity EO converter benefits from the increased single photon coupling rate ($g_{\mathrm{eo}}$) due to the reduction of resonator's modal volume, meanwhile providing high tunability and scalability towards multi-channel transduction and all-in-one devices. The compactness of integrated superconducting-photonic resonators, however, comes at the cost of resonator quality ($Q$) factor degradation and unintended microwave frequency shift, mainly arising from light-induced quasiparticle generation in superconductors \cite{xu2021bidirectional,fu2020ground,Holzgrafe:20}. 

In this letter, we present a study on the pulsed-light induced dynamics in the integrated superconducting EO transducer. An on-chip EO microwave-to-optical transducer on a thin-film lithium niobate (TFLN)-niobium nitride (NbN) hybrid material system is investigated as the experimental platform. Here, we observe a dynamic microwave frequency shifting in the superconducting cavity (``color'' change of microwave photons) resulting from its fast resonance frequency change induced by the optical pump pulse. The phenomena could be considered as an analog of the adiabatic frequency conversion in the optical domain, where the frequency of cavity photons is shifted due to a rapid change of the refractive index~\cite{yanik2005dynamic,notomi2006wavelength,preble2007changing,minet2020pockels,pang2021adiabatic}. These explorations reveal that optical illumination not only leads to static frequency variation and increased loss in the microwave resonator, it also imposes a fast, adiabatic microwave photon frequency shifting which must be considered in the design and calibration of microwave-to-optics converters. 
While this dynamic effect is regarded as deleterious for microwave-to-optics conversion, it could be exploited for "optically" control the microwave photon frequency at the cryogenic temperature or even shape the microwave-to-optical conversion process in the time domain. 

\section{Method}

\begin{figure}[t]
\centering
\includegraphics[width=0.48\textwidth]{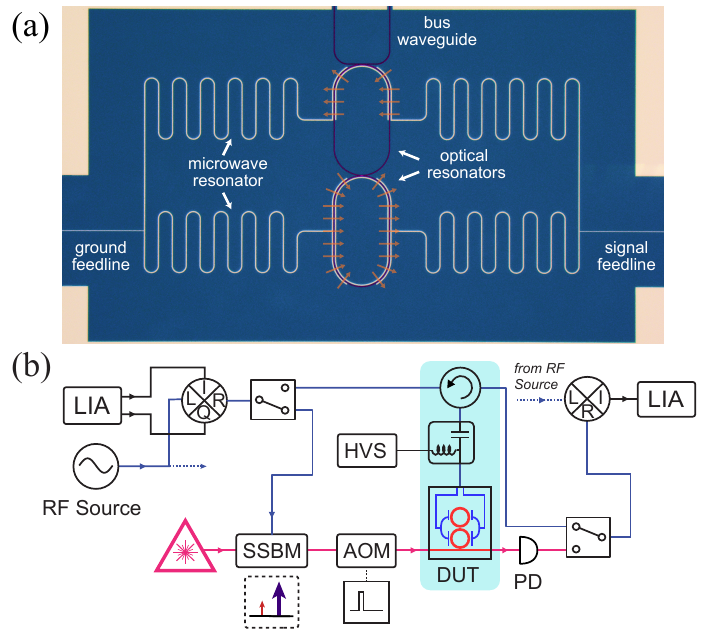}
\caption
{
(a) False colored microscopic image of the EO transducer device. Two strongly coupled optical racetrack resonators are integrated with a superconducting microwave resonator. The microwave resonator is linked to a coplanar waveguide (CPW, not shown) for both RF coupling and DC biasing. The electrical field direction of the designed microwave mode is marked in the coupling regime. (b) A schematic of the pulsed-pump measurement setup to characterize the response of the device in time domain. The microwave tone ($\sim$ \SI{6}{\giga\hertz}) is synthesized by up-converting a (\SI{50}{\mega\hertz}) intermediate frequency (IF) signal outputs from a lock-in amplifier (LIA) through an IQ mixer, while the optical signal is launched by optical single-sideband modulation (SSBM). The optical pump (together with optical signal) is pulsed via an acoustic-optic modulator (AOM) driven by a pulse generator. The microwave output signal is demodulated by the LIA after being down-converted back to \SI{50}{\mega\hertz}. PD: photodetector. HVS: high voltage source.}
\label{fig_1_schematic}
\end{figure}

The integrated EO transducer studied in this work consists of a superconducting microwave resonator and a pair of racetrack optical resonators, in which their electrical fields are coupled via Pockels nonlinearity of LN thin films \cite{tsang2010cavity,tsang2011cavity,javerzac2016chip} (Fig.\,\ref{fig_1_schematic}(a)). The two racetrack resonators, patterned from thin film lithium niobate, are strongly coupled to support a pair of hybridized transverse electric (TE) optical modes: anti-symmetric (mode $a$) and symmetric (mode $b$) \cite{soltani2017efficient}. The microwave-to-optical conversion efficiency is maximized when a triple resonance condition is satisfied by tuning their frequency spacing to match the microwave mode frequency via bias voltage. The superconducting resonator is made of NbN and supports a quasi-lumped LC resonant mode (mode $c$) that satisfies EO phase matching. 

Previously, it has been shown \cite{xu2021bidirectional} that the microwave Q of converter devices were limited by device packaging that introduces extra microwave loss through coupling to spurious modes. In this work, the microwave signal is coupled to the microwave resonator through a pair of on-chip feedline couplers connected to the coplanar waveguides (CPW). This permits the application of the DC bias voltage through the same set of couplers and avoids the use of separate DC electrodes used in our previous designs which could introduce extra microwave loss. The microwave cavity thus can be better shielded by both on-chip superconducting ground plates and an RF-tight copper box to eliminate the loss via coupling to spurious modes. With these improvements, we achieve an intrinsic microwave $Q_{\mathrm{in}}=17.1\mathrm{k}$, which is more than 10 times higher than that of the previous work.

The packaged device is loaded on the 1K plate ($T=$ \SI{800}{\milli\kelvin}) in a dilution fridge where high cooling power is available. The pulsed pump measurement setup is shown in Fig.\,\ref{fig_1_schematic}(b). We operate the transducer in a blue-drive (parametric amplification) scheme, in which the higher frequency optical mode $b$ is excited with a strong pump to stimulate the coherent coupling between optical mode $a$ and microwave mode $c$ \cite{PhysRevLett.109.130503,rueda2019electro}. 
Though this is not a quantum-state-conversion, it does not change the physics of the dynamic process we present in this work. A continuous wave (CW) tunable laser is tuned to the resonance wavelength (of mode $b$) and is pulsed via an \SI{80}{\mega\hertz} acoustic-optic modulator (AOM) with a pulse width of \SI{5}{\micro\second}. The optical pulses are sent to and collected from the device by a pair of grating couplers glued with angled lensed fiber \cite{mckenna2019alignment}. The total optical insertion loss is \SI{-22}{\deci\bel} within the fridge. To characterize the coherent response of the EO system in time domain, we utilize a lock-in-amplifier (LIA) as the signal generator and receiver. The microwave signal is synthesized from a \SI{50}{\mega\hertz} low frequency signal from the LIA via an IQ mixer, in which the up-converted frequency is tuned to match the microwave resonance. This GHz probe tone is then sent to the device either directly as the microwave input to the transducer, or as the optical input via optical single-sideband modulation (SSBM). The output microwave or optical signal from the device is then down-converted to \SI{50}{\mega\hertz} and sent back to LIA for quadrature measurement in time domain. 

\section{Result}

\begin{figure}[b]
\centering
\includegraphics[width=0.48\textwidth]{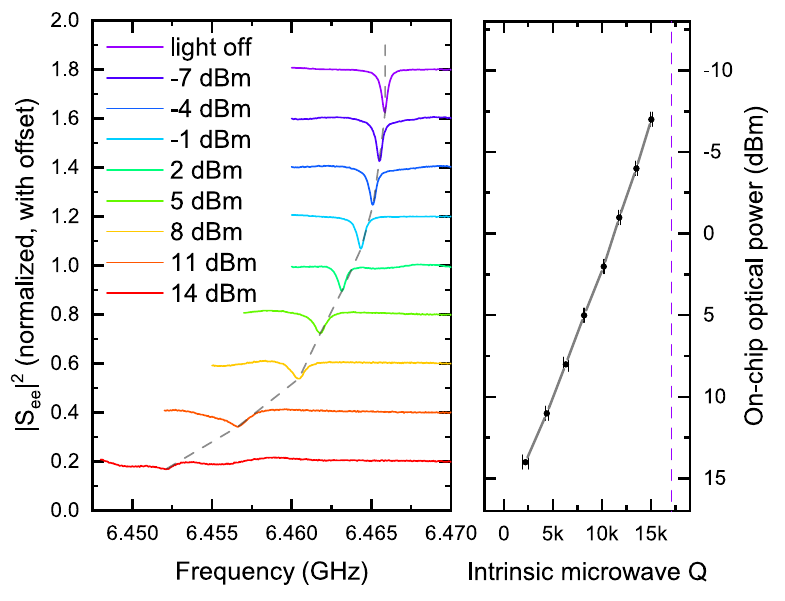}
\caption
{The microwave reflection spectra $|S_{\mathrm{ee}}|^2$ under different peak optical pump pulse power. The right panel shows the corresponding intrinsic microwave quality ($Q$) factor of the superconducting cavity. The no-light microwave $Q_{\mathrm{in}}=17.1\mathrm{k}$ is marked as purple dashed line.}
\label{fig_2_power}
\end{figure}

The dynamic process we study here arise from the fast photo-response of superconducting cavity which mainly originates from quasiparticle generation due to superconductor absorption of optical photons \cite{il2000picosecond,beck2011energy,uzawa2020optical}. When the optical pump pulse is turned on, the superconductor's absorption of light leads to a microwave resonance shift together with extra microwave loss. In Fig.\,\ref{fig_2_power}, we present the microwave reflection spectra $|S_{\mathrm{ee}}|^2$ under different peak optical power when the device reaches the steady state in the presence of the optical pump pulse. As the on-chip peak pump power is increased from \SI{-7}{\dBm} to \SI{14}{\dBm}, the microwave resonance exhibits a frequency shift from \SI{0.25}{\mega\hertz} to \SI{13.7}{\mega\hertz}, with its intrinsic $Q$ decreasing from 17.1\,k to 2.2\,k. For each trace shown in Fig.\,\ref{fig_2_power}, the repetition rate of the optical pulse is adjusted from \SI{4}{\kilo\hertz} to \SI{31.25}{\hertz} to keep the average power unchanged.

\begin{figure}[t]
\centering
\includegraphics[width=0.48\textwidth]{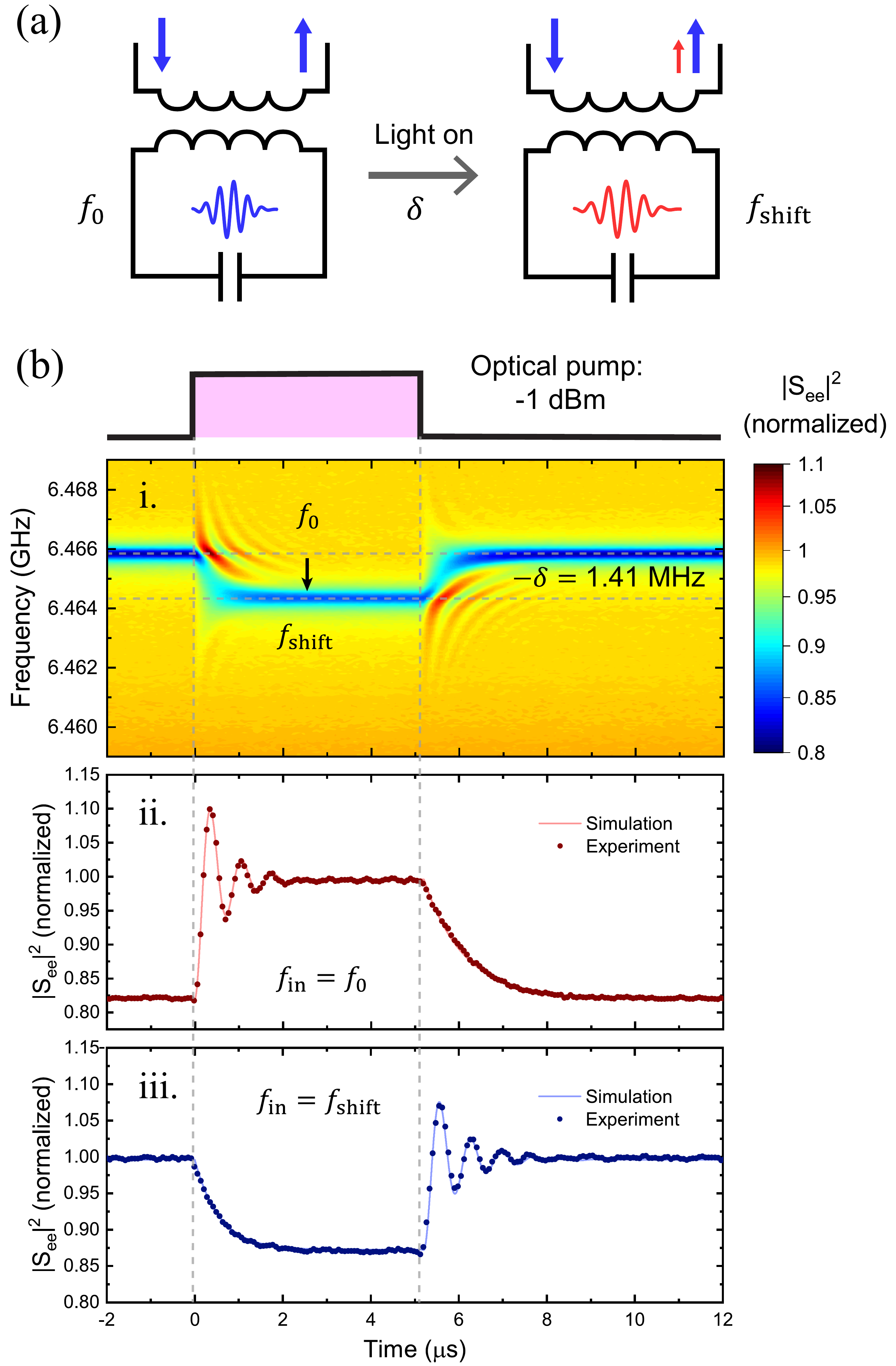}
\caption{(a) Illustration of light-induced dynamic frequency shifting in a superconducting cavity. When the light-induced resonance frequency change is faster than the cavity decay, the intracavity microwave photons is shifted to the new resonance frequency, resulting in a temporal interference fringe. (b) Experimental results exhibit the microwave dynamic frequency shifting. Panel (i) shows the time-resolved microwave reflection spectra $|S_{\mathrm{ee}}|^2$ for a pump pulse with \SI{-1}{\dBm} peak power. Panel (ii) and (iii) show the microwave reflection with input microwave frequency of $f_{0}$ (\SI{6.4658}{\giga\hertz}) and $f_{\mathrm{shift}}$ (\SI{6.4644}{\giga\hertz}), respectively. The microwave reflections exhibit an exponentially decaying oscillation when the optical pulse is on/off, indicating a beating between the shifted microwave signal and original input microwave signal.}
\label{fig_3_See}
\end{figure}

We then study the temporal response of the microwave cavity to the incident light by recording the microwave reflection spectrum at a fixed time delay with respect to the optical pulse. Here, we observe a dynamic frequency shifting of microwave photons in the superconducting cavity induced by the fast resonance frequency variation. A conceptional illustration of this dynamic frequency shifting process is shown in Fig.\,\ref{fig_3_See}(a). Before switching on the optical pulse, the CW input microwave signal with $f_{\mathrm{in}}=f_0$ is coupled on resonance to the microwave cavity. When the optical pulse arrives, the intracavity microwave photons frequency is rapidly modulated along the cavity mode to a new resonance frequency $f_{\mathrm{shift}}$. The beating between the emitted microwave cavity photons (at $f_{\mathrm{shift}}$)  and the reflected microwave signal (at $f_{\mathrm{in}}=f_0$) thus can be observed to verify this frequency shifting. This process can be considered as an analogy of the adiabatic frequency conversion in the optical domain, where the frequency of light traveling in a cavity is shifted by a fast variation in refractive index \cite{yanik2005dynamic,notomi2006wavelength,preble2007changing,minet2020pockels,pang2021adiabatic}. The requirement for this dynamic frequency shifting to be observable is that the cavity mode frequency shifts faster than the cavity photon lifetime. Because the photo-response time of thin-film NbN was reported to be less than \SI{1}{\nano\second} \cite{il2000picosecond,beck2011energy,uzawa2020optical}, the resonance frequency transition time is mainly defined by the optical pulse rise/fall time, which is estimated to be \SI{70}{\nano\second} in our system. Even with the lowest $Q=2.2$\,k, the corresponding cavity photon lifetime is $\tau_{\mathrm{e}}=$\SI {340}{\nano\second}, thus the requirement is satisfied.

\begin{figure}[b]
\centering
\includegraphics[width=0.42\textwidth]{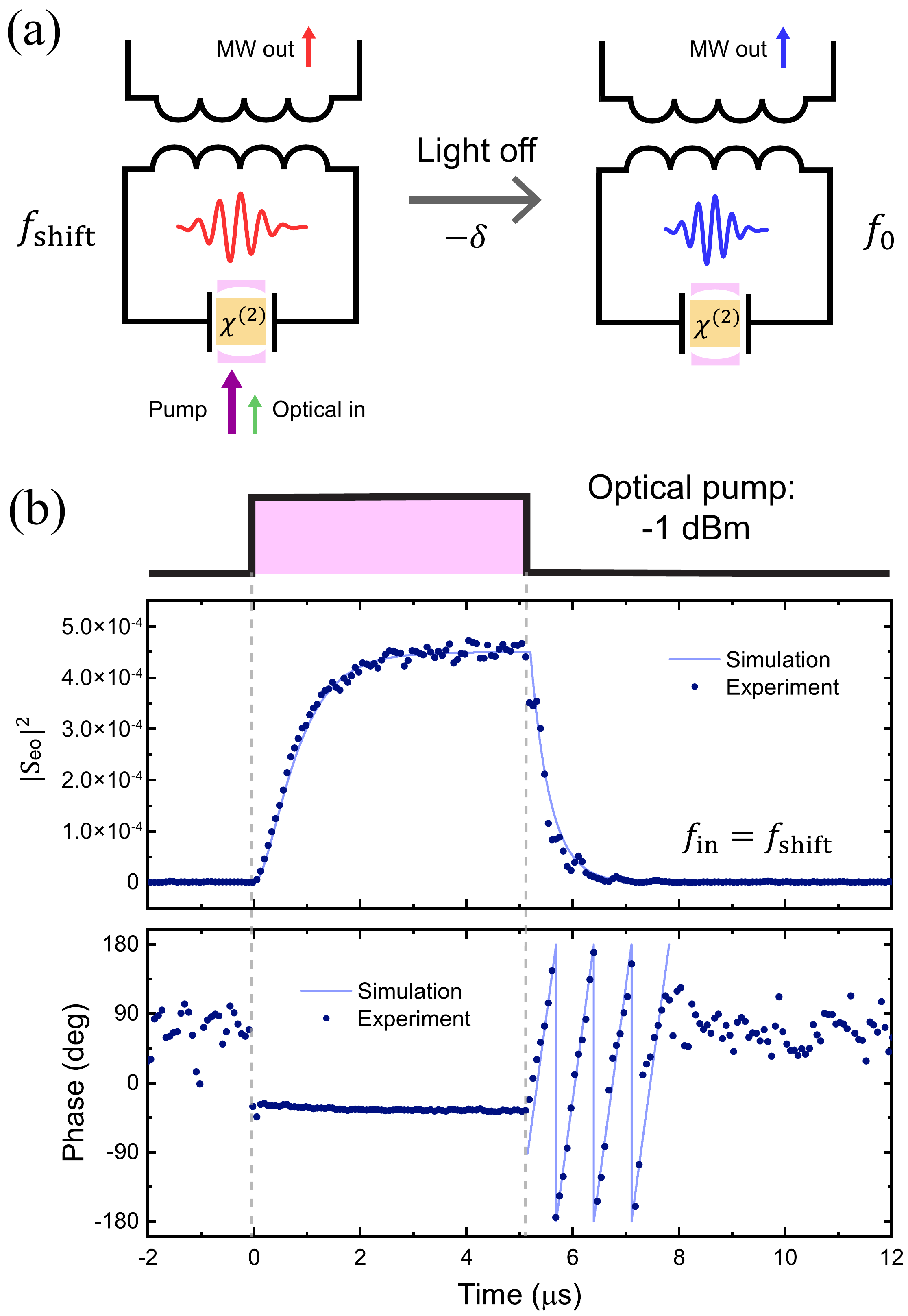}
\caption{ (a) An optical-to-microwave transduction cascade with a dynamic microwave frequency shift. (b) Time evolution of the optical-to-microwave scattering parameter $S_{\mathrm{eo}}$. The peak optical power is \SI{-1}{\dBm} and input frequency to optical SSBM is the shifted resonance $f_{\mathrm{shift}}=$ \SI{6.4644}{\giga\hertz}. When the pulse is turned off, the intracavity microwave photons converted from optical domain experience a secondary frequency shifting to the original ``un-illuminated'' resonance frequency $f_0=$ \SI{6.4658}{\giga\hertz}. A phase delay of $2\pi\delta t$ can be identified in the demodulated phase to confirm the frequency shifting, where $\delta =$ \SI{1.41}{\mega\hertz} is the change of resonance frequency.}
\label{fig_4_Seo}
\end{figure}

In Fig.\,\ref{fig_3_See}(b), we present experimental results that confirm this dynamic frequency shifting. Panel (i) presents the time evolution of the microwave reflection spectra $|S_{\mathrm{ee}}|^2$ when the device receives a \SI{5}{\micro\second}-long optical pump pulse with peak power of \SI{-1}{\dBm}. When the optical pump is turned on, the resonance frequency shift down from $f_0=$ \SI{6.4658}{\giga\hertz} by $\delta=$ \SI{1.41}{MHz} to $f_{\mathrm{shift}}=$ \SI{6.4644}{\giga\hertz}. The ripples in the time-resolved spectra indicate the interference between microwave signals with different frequency in the system. Specifically, with microwave input $f_{\mathrm{in}}=f_0$, the microwave reflection $|S_{\mathrm{ee}}|^2$ exhibits a decaying oscillation when the optical pump pulse is turned on, as shown in panel (ii). The decaying oscillation represents the interference between the frequency-shifted out-coupling microwave signal ($f_{\mathrm{shift}}$) and reflected input microwave signal ($f_0$), which is an evidence of the frequency shifting. Reversely, microwave signal with frequency $f_{\mathrm{in}}=f_{\mathrm{shift}}$ could be shifted back to $f_0$ when the the optical pump is switched off (panel (iii)). The oscillation period of \SI{0.71 }{\micro\second} corresponds to the frequency shift $\delta =$ \SI{1.41}{MHz}. The exponential decay rates of $|S_{\mathrm{ee}}|^2$ represent the different loss rates of microwave cavity when the optical pulse is turned on/off. The base level of $|S_{\mathrm{ee}}|^2$ 
reflects the extinction of the microwave resonance. We numerically simulate this dynamic process, and the results fit well with the experimental data. The details of the simulations are given in the Appendix. 

We next explore how this dynamic process affects the bidirectional microwave-to-optical transduction. In the pulsed transduction scheme, the microwave carrier is typically aligned with the shifted resonance frequency to optimize transduction efficiency. Figure\,\ref{fig_4_Seo}(b) shows time evolution of the optical-to-microwave scattering parameter $S_{\mathrm{eo}}$. The peak optical pump power is \SI{-1}{\dBm}, and the input frequency (to optical SSBM) is $f_{\mathrm{shift}}=$ \SI{6.4644}{\giga\hertz}. When the optical pulse is turned on, $|S_{\mathrm{eo}}|^2$ is gradually stabilized as the converted microwave photons build up in the superconducting cavity. Interestingly, we observe an optical-to-microwave transduction cascaded with a dynamic microwave frequency shifting here. An illustration of this cascaded process is shown in Fig.\,\ref{fig_4_Seo}(a). When the optical pulse is turned off, the intracavity microwave photons converted from optical domain experience another frequency shifting to $f_0=$ \SI{6.4658}{\giga\hertz}. When these new microwave photons ($f_0$) leak out from the cavity, there are no photons with frequency of $f_{\mathrm{shift}}$ in the cavity anymore, thus no oscillation behavior is observed here. Instead, by demodulating the output signal with the input frequency ($f_{\mathrm{shift}}=$ \SI{6.4644}{\giga\hertz}), the secondary dynamic shifting is confirmed by the phase delay with a period of \SI{0.71}{\micro\second} in the lower panel of Fig.\,\ref{fig_4_Seo}(b).

\begin{figure}[t]
\centering
\includegraphics[width=0.42\textwidth]{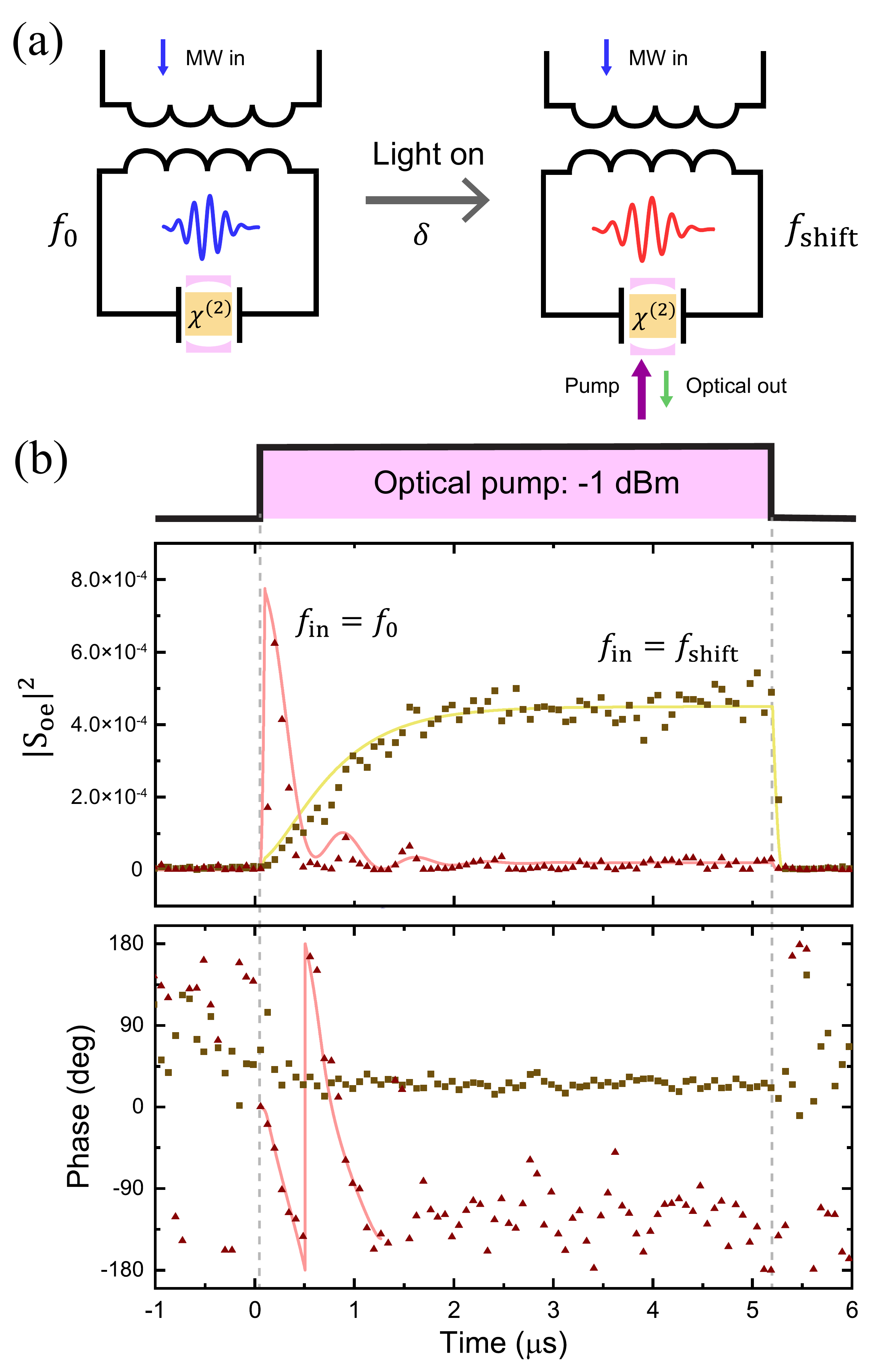}
\caption{ (a) A dynamic microwave frequency shifting cascade with a microwave-to-optical transduction. (b) Time evolution of the microwave-to-optical scattering parameter $S_{\mathrm{oe}}$ with peak pump power \SI{-1}{\dBm}. When $f_{\mathrm{in}}=f_0$, the preloaded microwave photons in the cavity are first shifted with the resonance frequency as the pulse turned on, and then are converted to optical photons. This cascaded processes exhibit a transient peak of $|S_{\mathrm{oe}}|^2$ and a phase delay in $S_{\mathrm{oe}}$.}
\label{fig_5_Soe}
\end{figure}

A similar cascade process can be further observed in a reverse manner, where the dynamic frequency shifting is followed with a microwave-to-optical transduction (Fig.\,\ref{fig_5_Soe}). The time evolution of the microwave-to-optical scattering parameter $S_{\mathrm{oe}}$ is shown in Fig.\,\ref{fig_5_Soe}(b). With the input microwave frequency aligned with the shifted resonance frequency $f_{\mathrm{shift}}$, $|S_{\mathrm{oe}}|^2$ is gradually stabilized as the microwave field builds up in the cavity (yellow trace). However, when the input microwave frequency is aligned with the native resonance frequency $f_0$, the cavity is pre-loaded by the CW microwave input before the pulse arrives. The stored microwave photons are first shifted with the microwave resonance as the pump pulse turned on, and then converted to optical domain. These cascaded processes result a transient peak together with a delayed phase in the demodulated $|S_{\mathrm{oe}}|$ scattering parameter. The pre-loaded cavity has a higher microwave $Q$ due to the absence of superconductor optical absorption. As a result, the instantaneous value of $|S_{\mathrm{oe}}|^2$ could be higher than the value in a stabilized transduction.


\section{Discussion and Conclusion}

The incident light introduces both static resonance frequency variation and dynamic photon frequency shifting. On the one hand, the light-induced resonance frequency variation, Q degradation, and added thermal noise excitations reported previously \cite{Holzgrafe:20,fu2020ground,xu2021bidirectional} are major limiting factors of the EO transducer device performance. Since scattered light from the chip-fiber interface is the dominant source of superconductor absorbed photons \cite{fu2020ground}, improving fiber-chip coupling and proper light shielding can be incorporated to mitigate this problem. On the other hand, the dynamic frequency shifting studied in this work pose another issue that must be considered and carefully engineered in the future microwave-to-optical transducer schemes. Nevertheless, this dynamic photon frequency shifting could be exploited, for example in the range where the microwave resonator is over coupled for bandwidth matching and the added noise by light is not a limiting factor. Meanwhile, how this dynamic photon frequency shifting interacts with the added microwave noise induced by the pump light would be interesting for further investigation.

The efficiency of this dynamic frequency shifting is determined by the microwave photon leakage during the resonance frequency transition time \cite{notomi2006wavelength,preble2007changing}. In the NbN superconducting resonator we used, the cavity microwave photon lifetime is much longer than the transition time, which is currently limited by the pulse synthesis timescale in the experiment. As a result, the intracavity efficiency is close to 100\%. However, for superconductors with a long quasiparticle lifetime, such as Aluminum \cite{PhysRevB.79.020509}, the resonance frequency transition time would be much longer, making this dynamic frequency shifting difficult to be observed. In our current device, the inductance part of the superconducting resonator have a wire width of \SI{4}{\micro\meter}. Under the same optical power, a much larger resonance frequency variation could be achieved in a lumped LC resonator with thinner wires and higher kinetic inductance~\cite{fu2020ground,niepce2019high}, thus the range of of this dynamic shifting can be expanded. 


In conclusion, the optical pulse-induced dynamics in integrated TFLN-superconductor hybrid platform are experimentally investigated. We describe and analyze a dynamic microwave frequency shifting in the superconducting cavity due to the fast resonance frequency variation induced by optical pump pulse. We further study how this "color change" of microwave photons plays a role in the bidirectional microwave-to-optical transduction. Our findings not only help understand the dynamics in the transducer, but also show a possible method to control the microwave photon frequency using optical pulses, which may be utilized to expand the transduction frequency range and thus to match the qubit excitation frequency.


\begin{acknowledgments}

This work is funded by ARO under grant number W911NF-18-1-0020. H.X.T. acknowledges partial supports from NSF (EFMA-1640959). Funding for substrate materials used in this research was provided by DOE/C2QA grant under award number DE-FOA-0002253. The authors thanks Michael Rooks, Yong Sun, Sean Rinehart and Kelly Woods for support in the cleanroom and assistance in device fabrication. 

\end{acknowledgments}

\appendix

\section{Theory of the EO system and device parameters}

The triply-resonant EO system is described by a Hamiltonian of \cite{tsang2010cavity,tsang2011cavity}
\begin{equation}
H=\hbar \omega_a a^{\dagger}a + \hbar \omega_b b^{\dagger}b + \hbar \omega_c c^{\dagger}c+g_{\mathrm{eo}}(ab^{\dagger}c+a^{\dagger}bc^{\dagger})
\end{equation}

where $a$, $b$ and $c$ denote annihilation operators for the lower (higher) frequency optical modes and microwave mode respectively. The resonance frequency of $a$, $b$, $c$ are $\omega_a$, $\omega_b$, $\omega_c$ respectively, and $g_{\mathrm{eo}}$ denotes the single photon electro-optical coupling rate. In the parametric amplification scheme, the mode $b$ is coherently driven with a strong optical pump, and the system Hamiltonian in the rotation frame of the optical pump can be simplified to \cite{fan2018superconducting,rueda2019electro}
\begin{equation}
H=\hbar \delta_a a^{\dagger}a + \hbar \omega_c c^{\dagger}c +\hbar G (ac+a^{\dagger}c^{\dagger})    
\end{equation}
where $\delta_a=\omega_a-\omega_p$ and $G=\sqrt{n_p}g_{\mathrm{eo}}$ is pump photon number ($n_p$) enhanced coupling rate. Thus the equations of motion could be written as 
\begin{align}
\label{eq1}
& \frac{d}{dt}a=-(i\delta_a+\frac{\kappa_a}{2})a-iGc^*+\sqrt{\kappa_{a,\mathrm{ex}}} a_{\mathrm{in}} e^{-i\delta_{a,\mathrm{in}}t} \\
\label{eq2}
& \frac{d}{dt}c=-(i\omega_c+\frac{\kappa_c}{2})c-iGa^*+\sqrt{\kappa_{c,\mathrm{ex}}} c_{\mathrm{in}} e^{-i\omega_{c,\mathrm{in}}t} \\
\label{eq3}
& a_{\mathrm{out}}=a_{\mathrm{in}}-\sqrt{\kappa_{a,\mathrm{ex}}} a \\
\label{eq4}
& c_{\mathrm{out}}=c_{\mathrm{in}}-\sqrt{\kappa_{c,\mathrm{ex}}} c
\end{align}
where $\kappa_{a}$, $\kappa_{a,\mathrm{ex}}$ ($\kappa_{c}$,$\kappa_{c,\mathrm{ex}}$) are the total decay and external coupling rate for mode $a$ ($c$) respectively. $a_{\mathrm{in}}$ ($a_{\mathrm{out}}$) and $c_{\mathrm{in}}$ ($c_{\mathrm{out}}$) denote the input (output) signals, $\delta_{a,\mathrm{in}}$ and $\omega_{c,\mathrm{in}}$ are the angular frequency of the inputs. The dynamic of the system could be numerically simulated in the time domain with Eqs.~(\ref{eq1})-(\ref{eq4}). In our simulation model, we simply assume $\omega_c$ linearly change to the final frequency within a transition time $\tau$,

\begin{equation}
\omega_c(t)= \left\{
\begin{aligned}
& \omega_0+\frac{t-t_1}{\tau}(\omega_{\mathrm{shift}}-\omega_0) & & t_1<t<t_1+\tau \\
& \omega_{\mathrm{shift}}-\frac{t-t_2}{\tau}(\omega_{\mathrm{shift}}-\omega_0) & & t_2<t<t_2+\tau
\end{aligned}
\right.
\end{equation}
where $\omega_0$ and $\omega_{\mathrm{shift}}$ are the light-off and light-on microwave angular frequency, respectively. $t_1$ and $t_2$ are the pulse-on and pulse-off time, respectively; and $\tau$ is the optical pulse rise/fall time.  A more complex model including real nonlinear terms is worth further investigation with more detailed knowledge of the material properties.

By solving the Eqs.~(\ref{eq1})-(\ref{eq4}) in steady state, we can get the full scattering matrix of the transducer system. When all resonances are perfectly aligned ($\omega_a+\omega_c=\omega_b$) and $C<1$, the on-chip transducer scattering parameter is 
\begin{equation}
\label{eq5}
  |S_{\mathrm{eo}}|^2=|S_{\mathrm{oe}}|^2=\frac{\kappa_{a,\mathrm{ex}}}{\kappa_{a}}\frac{\kappa_{c,\mathrm{ex}}}{\kappa_{c}}\frac{4C}{(1-C)^{2}},
\end{equation}
where figure of merit of this parametric amplification process is the cooperativity
$C=4 n_\mathrm{p} g_{\mathrm{eo}}^2 / \kappa_a \kappa_c$.
When $C \geq 1$, the system should be considered with a pump depletion, and enters a parametric oscillation regime. 

The parameters of the device is shown in Table.\ref{table_1}. The on-chip transduction efficiency of the device is also calibrated by the full spectra of the complete scattering matrix \cite{andrews2014bidirectional}, and the cooperativity $C$ is extracted from Eq.~(\ref{eq5}) accordingly. Here, a cooperativity of $C=0.107 \pm 0.06$ is reached in this device with on-chip \SI{5}{\dBm} pump power, which is more than twice higher than our previous result under \SI{8}{\deci\bel} weaker pump power.

 \begin{table}[htb]
    \centering
    \renewcommand{\arraystretch}{1.2}
    \begin{tabular}{| m{2.2cm}<{\centering} |  m{3cm}<{\centering} |  m{2.5cm}<{\centering} |}
    \hline
        Parameter & Description & Value  \\
    \hline
    $(\omega_a , \omega_b)/2\pi$ \quad \quad \quad \quad \quad& optical resonance frequency & (191.698, 191.704)\,THz \\
    \hline
    $(\kappa_a , \kappa_b)/2\pi$ & total optical loss rate & (197, 238)\,MHz \\ 
    \hline
    $(\kappa_{a,\mathrm{ex}} , \kappa_{b,\mathrm{ex}})/2\pi$ \quad \quad \quad \quad \quad &external optical coupling rate & (13, 65)\,MHz \quad \quad \quad \quad \quad \quad \quad \quad \quad \quad \\ 
    \hline
    $(\kappa_{a,\mathrm{in}} , \kappa_{b,\mathrm{in}})/2\pi$ \quad \quad \quad \quad \quad \quad \quad &intrinsic optical loss rate & (185, 173)\,MHz \quad \quad \quad \quad \quad \quad \quad \quad \quad \quad \\ 
    \hline
    $\omega_c/2\pi$ \quad \quad \quad \quad \quad \quad \quad \quad \quad \quad & microwave resonance frequency & 6.4658\,GHz \quad \quad \quad \quad \quad \quad \quad \quad \quad \quad \\
    \hline
    $\kappa_c/2\pi$ \quad \quad \quad \quad \quad \quad \quad \quad \quad \quad & total microwave loss rate & 397\,kHz \quad \quad \quad \quad \quad \quad \quad \quad \quad \quad \\
    \hline
    $\kappa_{c,\mathrm{ex}}/2\pi$ \quad \quad \quad \quad \quad \quad \quad \quad \quad \quad & external microwave coupling rate & 19\,kHz \quad \quad \quad \quad \quad \quad \quad \quad \quad \quad \\
    \hline
    $\kappa_{c,\mathrm{in}}/2\pi$ \quad \quad \quad \quad \quad \quad \quad \quad \quad \quad & intrinsic microwave loss rate & 378\,kHz \quad \quad \quad \quad \quad \quad \quad \quad \quad \quad \\
    \hline
    $g_{\mathrm{eo}}/2\pi$ \quad \quad \quad \quad \quad \quad \quad \quad \quad \quad & single photon EO coupling rate & 500\,Hz \quad \quad \quad \quad \quad \quad \quad \quad \quad \quad \\
    \hline
    \end{tabular}
    \caption{
    \label{table_1} \textbf{Device parameters.} Note that the parameters of the microwave resonator is calibrated in the absence of optical light. The decrease of intrinsic microwave $Q$ is shown in Fig.\,\ref{fig_2_power}, while the external coupling rate only slightly change under different optical power. }
\end{table}

\end{document}